\documentclass[aps,prl,twocolumn,showpacs,superscriptaddress]{revtex4-1}
\usepackage{graphicx}
\usepackage{subfigure}
\usepackage{epstopdf}
\usepackage{amsmath}
\usepackage{amssymb}
\usepackage{float}
\usepackage{bm}
\usepackage{mathtools}
\usepackage{multirow}
\usepackage{color}

\input{epsf}
\begin{document}

\title{Few-body collective excitations beyond Kohn's theorem in quantum Hall systems}

\author{R.~E. Wooten}
\affiliation{Department of Physics and Astronomy,
Purdue University,
West Lafayette, Indiana 47907, USA}
\author{B. Yan}
\affiliation{Department of Physics and Astronomy,
Purdue University,
West Lafayette, Indiana 47907, USA}
\author{Chris H. Greene}
\affiliation{Department of Physics and Astronomy,
Purdue University,
West Lafayette, Indiana 47907, USA}
\affiliation{Purdue Quantum Center, Purdue University, West Lafayette, Indiana 47907, USA}
\date{\today}

\begin{abstract}

A relative coordinate breathing mode in the quantum Hall system is predicted to exist with different behavior under either Coulomb or dipole-dipole interactions.  
While Kohn's theorem~\cite{Kohn1961} predicts that any relative coordinate interaction will fail to alter the center of mass energy spectrum, it can affect excitations in the relative coordinates.
One such collective excitation, which we call the hyperradial breathing mode, emerges naturally from a few-body, hyperspherical representation of the problem and depends on the inter-particle interactions, the ground state wave function, and the number of particles participating in the excitation. 
Possible observations of this excitation will be discussed in the context of both cold rotating atomic simulations and conventional quantum Hall experiments.

\end{abstract}


\maketitle

In condensed matter and atomic physics alike, particle interactions can give rise to collective behaviors in many-body systems with dramatic and often unexpected properties~\cite{Bardeen1957, Osheroff1972, Tsui1982, Davis1995}.
In no system are collective behaviors more central than in the fractional quantum Hall (QH) system, where the properties of low energy quasiparticle excitations continue to drive new theoretical and experimental discoveries~\cite{Laughlin1983,Haldane1983,Jain1989,Moore1991,Son2015, Samkharadze2016}.
However, collective excitations at higher energies, near the cyclotron frequency $\omega_c = eB/m$, are considerably less well explored: while the the center of mass excitation, which is indistinguishable in frequency from the single particle excitation frequency $\omega_c$ by Kohn's theorem~\cite{Kohn1961}, is clearly predicted by theory, experiments in capacitance spectroscopy~\cite{Ilani2004}, optical emission spectroscopy~\cite{Levy2016}, and high-intensity pulsed terahertz spectroscopy~\cite{Maag2016} have detected behaviors that defy the simple single-particle or Kohn's theorem predictions, indicating that the cyclotron frequency excitation regime exhibits interesting new physics.
A variety of numerical treatments have been used to characterize QH systems~\cite{Yoshioka1983,Haldane1985,Jeon2004,Nishi2006,Yannouleas2011}, however isolating any specific excitation in the cyclotron energy regime from many of these models is daunting since the excitation spectrum is highly complicated.

However, recasting the quantum Hall (QH) problem in the adiabatic hyperspherical representation~\cite{Daily2015,Rittenhouse2015} highlights the existence of a unique type of vibrational mode~\cite{Bao1996} that may be directly measurable. 
The adiabatic hyperspherical representation~\cite{Macek1968JPB,fano1981,fano1983} has not seen widespread use in condensed matter physics, but has contributed to developments in many disparate fields, including 
nuclear structure and reactivity~\cite{smirnov1977Sov.J.Part.Nucl.,avery1989, avery1993JPC, nielsen2001PRep, DailyKievskyGreene2015},
universal Efimov physics in cold atoms and 
molecules~\cite{WangDIncaoEsry2013,Rittenhouse2011JPB, Wang2014arXiv1412p8094,efimov1970PLB, nielsen1999PRLb,gattobigio2012PRA},
few-electron atoms~\cite{lin1986AMOP,lin1995PRep, lin2000PHYSICSESSAYS}, 
positron and electron systems~\cite{botero1985PRA,DailyGreene2014,DailyGreene2015PRA,archer1990PRA},
Bose-Einstein condensates~\cite{BohnEsryGreene1998PRA,kushibe2004PRA}, 
and trapped 
degenerate Fermi gases~\cite{rittenhouse2006PRA,Rittenhouse2011JPB}. 
The technique is broadly useful because it expresses few- or many-body interacting systems in collective coordinates and separates a particle cluster's internal geometry from the cluster center-of-mass motion. 
This Letter examines the origin and properties of a particular type of vibrational excitation observable in quantum Hall systems, which we call the hyperradial breathing mode, and also discusses possible schemes for its measurement in experiments in both condensed matter and cold atom systems.

Consider the many-body QH Hamiltonian for $N$ electrons confined to two-dimensions in a strong, perpendicular magnetic field in the symmetric gauge~\cite{jainbook}.
A rotating two-dimensional gas of neutral atoms in a harmonic trap (or even non-rotating, see below) shares the same Hamiltonian, except for the form of the interactions~\cite{Cooper2008}, making it an ideal system for comparing the effect of different interactions on the collective behaviors of the system.
The hyperspherical transformation first extracts the center of mass coordinate from the Hamiltonian then converts the remaining $2N-2$ relative Jacobi coordinates into $2N-3$ angular dimensions known as hyperangles, collectively labeled $\Omega$, and a single length scale known as the hyperradius, $R$.
The hyperradius is a scalar whose square is equal to the sum of the squares of the mass-weighted relative Jacobi coordinates and essentially defines an approximate area covered by the $N$-particle system.
All lengths in this Letter have been scaled by the magnetic length, $\lambda_0 = \sqrt{\hbar/m_e \omega_c}$ for the condensed matter system, where $m_e$ is the effective mass of the electron in the material, or by the trap length in the cold atom system, $\lambda_0 = \sqrt{\hbar/m \omega}$, where $m$ is the atom mass and $\omega$ is defined as twice the trap frequency~\cite{Cooper2008}, and is the analog of the Landau level spacing in the conventional system.
For brevity, "$\hbar \omega$" will be used to represent the Landau level separation in both systems when appropriate.
 
In the symmetric gauge, the relative coordinate interacting Hamiltonian can be rewritten in hyperspherical coordinates as
\begin{align}
  H_{\rm rel} = & -\frac{1}{2\mu} {\bm \nabla}_{R,\Omega}^2 + \frac{\mu}{8}R^2 + \frac{1}{2}L_z^{\rm rel} + \kappa C({\bf \Omega})V(R),  \label{eq.1}
\end{align}
where ${\bf \nabla}_{R,\Omega}^2$ is the Laplacian in hyperspherical coordinates \cite{avery1989}, $\mu = N^{-1/N_{rel}}$ is a dimensionless mass scaling factor, and $C(\bm{\Omega})$ is the hyperangular part of the interactions.
The last term in Eq.~\eqref{eq.1} represents the interactions in terms of the hyperspherical coordinates, with the lengths scaled by $\lambda_0$ for the system in question.
For the condensed matter system, the interactions are simply Coulomb repulsive, 
but in two-dimensional cold atom systems, a variety of interactions can be implemented by different experimental choices. We restrict our cold atom investigation to clusters of electric or magnetic dipoles aligned with the axis of rotation interacting purely via repulsive dipole-dipole interactions, which are among the class of interactions that can drive the formation of quantum Hall liquids~\cite{ Quinn2000,Wojs2001,Osterloh2007}. 
Then the term $\kappa$ in Eq.~\eqref{eq.1} is the ratio of the interaction energy to the Landau level separation: 
for Coulomb interactions, 
$\kappa = e^2/(4 \pi \epsilon \lambda_0 \hbar \omega_c)$;
for dipole-dipole interactions,
$\kappa = c_{dd}/(4 \pi \lambda_0^3 \hbar \omega)$, where $c_{dd}= \mu_0 \mu_{mag}^2$ for polarized magnetic dipoles with magnetic moment $\mu_{mag}$, and $c_{dd}= d^2/ \epsilon_0$ for polarized electric dipoles with dipole moment $d$.
The form of the hyperangular term $C({\bf \Omega})$ depends on the form of the interactions and on the specific choices of Jacobi vectors and hyperspherical coordinates, and $V(R)$ takes the simple forms $1/R$ for Coulomb interactions or $1/R^3$ for polarized dipole-dipole interactions.
\begin{figure}
  \centering
  \includegraphics[angle=0,width=0.48\textwidth]{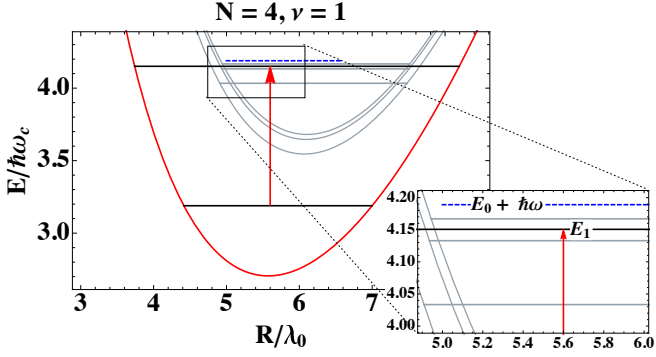} 
  \caption{(Color online) 
  Hyperradial potential curves and the hyperradial bound states for the four particle, $\nu = 1$ system.  The ground state for $\nu = 1$ has hyperangular quantum numbers $K = 6, M = -6$ and $n_R=0$, and is totally isolated. The hyperradial excitation takes the system from the $n_R=0$ state with energy $E_0$ to the first hyperradially excited state with $n_R=1$ and energy $E_1$. The $E_0$ plus the cyclotron energy is shown as a (blue) dashed line for contrast. Other excited potential curves and their ground state energies with the same $M$ with are shown in pale grey.
   }
  \label{fig1}
\end{figure}
In the absence of interactions, the quantum Hall Hamiltonian is exactly separable into a hyperradial and a hyperangular Hamiltonian. 
The solutions are products of hyperradial functions times hyperangular functions known as the hyperspherical harmonics from K-harmonic theory~\cite{smirnov1977Sov.J.Part.Nucl.}, $\Psi(R,\Omega) = R^{-N_{rel}+1/2}F_{n_R,K}^{(M)}(R)\Phi^{(M)}_{K,i}({\bf \Omega})$.  
The quantum number $M$ is the familiar total relative azimuthal quantum number which determines the filling factor $\nu$, $K$ is an additional hyperangular momentum quantum number known as the grand angular momentum, and $n_R = 0, 1, \ldots$ is a hyperradial nodal quantum number.
Particle exchange symmetry of the basis functions is imposed in a separate step, for example by a technique developed by Efros~\cite{efros1995} as outlined in \cite{Daily2015}.

Diagonalizing the fixed-$R$ hyperangular Hamiltonian in a restricted hyperangular Hilbert space (a reasonable approximation~\cite{Daily2015} is to ignore coupling between different K-manifolds; for $K = |M|$, this coincides with restricting the Hilbert space to the lowest Landau level) in degenerate perturbation theory with $R$ as an adiabatic coordinate reduces the many-dimensional hyperradial Sch\"odinger equation to a set of one-dimensional, uncoupled, ordinary differential equations:
\begin{equation}
\left\{ -\frac{1}{2\mu} \frac{d^2}{dR^2} + U_{K,a}^{(M)}(R) - E \right\} F_{n_R,K,a}^{(M)}(R) = 0, \label{eq.2}
\end{equation}
where the $U_{K,a}(R)$ are the hyperradial potential curves, and $a$ is a label to distinguish different curves in the same $K,M$ manifold. 
In the noninteracting limit, the potential curves take the form
\begin{align}
  \label{eq_UNI}
  U_{K}^{(M)}&(R) =  \\ \nonumber
  &\frac{(K+N_{\rm rel}-1/2)(K+N_{\rm rel}-3/2)}{2\mu R^2}
  + \frac{\mu}{8}R^2 + \frac{1}{2}M.
\end{align}
Each of these potential curves $U_{K,a}^{(M)}(R)$ supports a collection of energies separated by $\hbar \omega$ zero-th order in $\kappa$,   $E^{0} = (2n_R + M + K + N_{\rm rel})\hbar \omega/2$. 
When the $\kappa$-dependent term in the adiabatic potential curve is treated in first order perturbation theory, the energies for the Coulomb system exactly match those calculated in conventional configuration interaction calculations~\cite{Daily2015,Jeon2007}. If the adiabatic hyperradial differential equation is solved numerically, Ref.~\cite{Daily2015} shows that even higher accuracy is obtained.

Calculations beyond first-order perturbation theory (e.g. using finite differences techniques to solve Eq.~\eqref{eq.2}), reveal that interaction induced shifts to the potential curves cause the hyperradial excitations within each curve to shift so their separations no longer exactly equal $\hbar \omega$. 
As a simple example, Fig.~\ref{fig1} shows the hyperradial curves and energies for the four-particle integer quantum Hall state ($M=-6$, $K=6$, and $n_R=0$) in GaAs and all excited states with $M=-6$ that are approximately $\hbar \omega_c$ higher in energy.
We are interested in the lowest energy hyperradial excitation, the transition from $n_R = 0$ to $n_R = 1$ with energy $(E_1 - E_0)$ for any given set of hyperangular quantum numbers, as is highlighted with the vertical (red) arrow in Fig.~\ref{fig1}.
Using more exact numerical techniques here constitutes including some level of Landau level mixing in our approximation; our previous studies give bounds to the hyperangular contribution to Landau level mixing, and indicate that hyperangular Landau level mixing, or coupling between K-manifolds, is weak for lowest Landau level ground states and modest values of $\kappa$.

\begin{figure*}
  \centering
  \includegraphics[angle=0,width=0.9\textwidth]{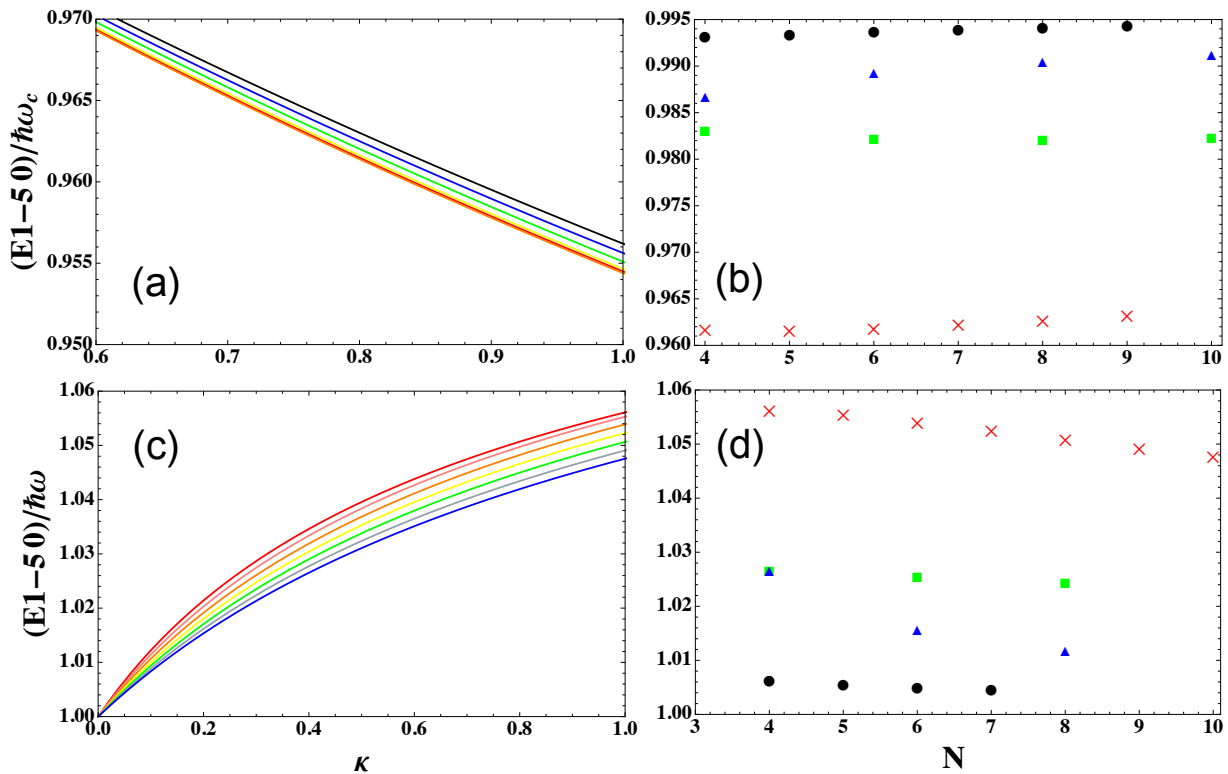} 
  \caption{(Color online)
  (a) The hyperradial vibrational (HRV) mode excitation energies for Coulomb interactions as a function of $\kappa$ for $N=4,\ldots,9$ particles at $\nu=1$ filling factor. From the lowest curve (red) with $N=4$, the number of particles increases to $N=9$ for the uppermost (black) curve.  (b) The HRV energies for Coulomb interactions versus particle number for various filling factors [$\nu=1$ (red 'x's), $\nu=2/3$ (green squares), $\nu=2/5$ (blue triangles), and $\nu=1/3$ (black circles)]. The values of $\kappa$ are calculated from corresponding experimental magnetic fields in Tesla from \cite{Eisenstein1990}: $\nu=1$ corresponds to $9T$, $\nu=2/3$ to $14T$, $\nu = 2/5$ to $25T$, and $\nu=1/3$ to $29T$. (c) The HRV excitation energies for dipole-dipole interactions as a function of $\kappa$. In this case the number of particles decreases from $N=4$ downward to $N=9$ on the plot.  (d) The HRV energies for dipole-dipole interactions at $\kappa=1$. The filling factors are labeled as in (b).
 }
\label{fig2}
\end{figure*}
Fig.~\ref{fig2} gives the energy separation between the ground state and the first hyperradial excited state for several important lowest Landau level filling factors in the GaAs system (top) and the cold atom system (bottom) as a function of $\kappa$ (left) and the number of particles (right).
For Fig.~\ref{fig2}(b), the values of $\kappa$ used for each filling factor are taken from the experimental results of \cite{Eisenstein1990}. 
Since cold atom systems are hypothetically more tunable and currently lack an experimental paradigm, $\kappa$ was set to $1$ for all filling factors of the dipole-dipole interaction calculations shown in Fig.~\ref{fig2}(d).
We note that the hyperradial excitation energy $E_1-E_0$ for Coulomb repulsion is smaller than $\hbar \omega_c$ for all tested systems, while the opposite is true for the dipole-dipole interacting system.
In general, the vibrational mode excitation energy detuning from $\hbar \omega$ in both cases is largest when $\kappa$ is large and when the filling factor is smallest.
Increasing the number of particles weakens the detuning in both cases as well, although this trend does not hold universally for Coulomb interactions, as there are a few exceptions (which are difficult to see in the scales of Fig.~\ref{fig2}(b)).
We have not yet found a simple interpretation for the non-monotonic N-dependence of the energy shifts.
Overall, the effect of dipole-dipole interactions is weaker than the effect of Coulomb interactions, and the more dramatic detuning due to Coulomb interactions is due to the Coulomb interaction's longer-range nature.

Experimentally, this hyperradial breathing mode cannot be excited through purely optical means because the laser field only operates on the center-of-mass for equal-mass, equal-charge particles, and should not induce transitions of the internal degrees of freedom of the system without additional terms in the Hamiltonian involving significant coupling between the center-of-mass and relative degrees of freedom (which we neglect in this work, but could include localized anisotropic features of the background, e.g. impurities or lattice defects).
However, the transition could be induced by a time dependent perturbation to the radial harmonic confinement of the form
\begin{equation}
V'(t) = a \cos{\omega_0 t} \sum_{i=1}^N r_i^2 = a \cos{\omega_0 t} \left( \frac{1}{N} r_{cm}^2 + \mu R^2 \right), 
\label{eq.3}
\end{equation}
where $a$ is the strength of the weak potential, $r_i$ are the single-particle coordinates, and $\omega_0$ is the hyperradial transition frequency. 
From the form of Eq.~\eqref{eq.3}, it is clear that such a potential can perturbatively excite the center of mass or the hyperradial degrees of freedom, but the hyperradial excitation can be spectroscopically selected by the choice of frequency.
For a two-dimensional electron gas, this oscillating potential could be achieved by weakly oscillating the perpendicular magnetic field at high frequencies, although the terahertz frequencies required for typical samples will be experimentally challenging to achieve, and detection will also prove difficult.

Measuring the hyperradial breathing modes should be more feasible in trapped cold atom or cold molecule systems interacting via repulsive dipole-dipole interactions, where the harmonic perturbation of Eq.~\eqref{eq.3} can be produced by flexing the trapping potential in a time dependent manner.
Collective modes have been previously observed directly in Bose-Einstein condensates~\cite{Jin1996} and degenerate Fermi gases, including in two-dimensions~\cite{Vogt2012}, through such techniques in the absence of internal rotation as a tool to evaluate various internal properties of the gas.


Construction of a cold atom or cold molecule quantum Hall gas remains a significant experimental challenge, but the cold atom systems present a dramatic range of tunability which could be ideal for probing these vibrational modes.
Using, for example, the magnetic dipole interactions of $^{161}Dy$~\cite{Lu2012} and assuming a $\nu=1$ filling factor, dipole trap with a planar trapping frequency of $\omega_{\rm trap}=30$kHz has $\kappa$ of only $2.6\times10^{-3}$, and a detuning of only $\approx 22$Hz, but a much tighter trap could enhance the detuning, since $\kappa$ varies with the square root of $\omega$. 
Substituting magnetic dipolar atoms with cold electric-dipolar bialkali molecules, which have intrinsic dipole moments of around 1 Debye, can also dramatically enhance the effect.
For example, fermionic LiRb has an intrinsic dipole moment of around $4.1$ Debye~\cite{Aymar2005}, so in a 15kHz trap, $\kappa \approx 1.5 $ and the detuning for the $N=4, M=6$ integer quantum Hall state is around 2 kHz.
The greatest challenge in this experimental system will be the measurement of the energy, but there are several methods that might prove effective.
For an array of quantum Hall droplets, photoassociation measurements in the spirit of~\cite{Baur2008,Gemelke2010} may be sensitive enough to measure the few-body excitation energies.
Alternately, it may be feasible to directly measure the total absorption of the perturbative light by the many droplets.
If instead the successful quantum Hall experiment consists of a single droplet of only a few particles in a deep-well, optical tweezer, the excitation energy might be measured by Coulomb explosion imaging~\cite{Kunitski2015} or by a sensitive trap loss~\cite{Serwane2011}.  

While most discussions of the quantum Hall effect for ultracold atoms have envisioned rotating traps, it should be pointed out that the spectra predicted here can be observed also for a nonrotating isotropic 2D trap.  This is because the difference in the Hamiltonian between a rotating versus a non-rotating trap is simply the presence of the constant term $\frac{1}{2}L_z^{\rm rel}$ in Eq.~\eqref{eq.1}, which is present only for a rotated trap in the rotating frame.  But since $L_z^{\rm rel}$ is a conserved quantity for this system, the energy levels should be observable if the appropriate relative angular momentum modes are created for the number of atoms or molecules in the trap. For example, in a non-rotating 2D trap containing 4 identical, spin-polarized fermionic atoms, the Laughlin $\frac{1}{3}$ state is the lowest energy eigenstate having relative angular momentum $|M|=18$, and the breathing mode frequency predicted using the adiabatic hyperspherical approximation should be accurate.  It is therefore an observable excitation in the Hilbert space even though it is not the $M$ value of the ground state of the system as a whole.

In conclusion, we have established the existence of a hyperradial breathing mode in the quantum Hall system. 
This breathing mode energy is affected by the particle count, the strength and type of the interaction, and the filling factor.
Although experimental realizations of this measurement face significant challenges, the modes should be experimentally excitable and measurable.
As a final speculation, we suggest that, while exciting in its own right as a new collective excitation in the system, this particular measurement could also be useful in establishing direct measurements of the effects of Landau level mixing (deviations from the ideal hyperradial vibrational mode energies can be attributed to hyperangular coupling between Landau levels), which have been estimated through various methods~\cite{Simon2013,Sodemann2013,Peterson2013,Wooten2013}, but are challenging to measure experimentally.



\section{Acknowledgments}

This work has been supported in part by a Purdue University
Research Incentive Grant from the Office of the Vice President for Research. Some numerical calculations were performed under NSF XSEDE Resource Allocation No. TG-PHY150003. We also thank Kevin Daily for access to unpublished computational codes used in this work.

\bibliography{mybib}{}

\end{document}